\documentclass[12pt]{article}
\usepackage{eurosym}
\usepackage{graphicx}
\usepackage{mathptmx,mathrsfs}
\usepackage{amsfonts,amsmath,amssymb,amsthm}
\usepackage{indentfirst}
\usepackage{cite}
\usepackage{hyperref}

\numberwithin{equation}{section}

\theoremstyle{definition}

\begin{document}

\title{Lorentz-violating effects in three-dimensional $QED$}
\author{R. Bufalo$^{1}$\thanks{%
rbufalo@ift.unesp.br}~ \\
\textit{{\small $^{1}$ Instituto de F\'{\i}sica Te\'orica (IFT), UNESP,
S\~ao Paulo State University} }\\
\textit{\ {\small Rua Dr. Bento Teobaldo Ferraz 271, Bloco II Barra Funda,
CEP 01140-070 S\~ao Paulo, SP, Brazil}}\\
}
\maketitle
\date{}

\begin{abstract}
Inspired in discussions presented lately regarding Lorentz-violating
interaction terms in \cite{13,6}, we propose here a slightly different version
for the coupling term. We will consider a modified quantum electrodynamics
with violation of Lorentz symmetry defined in a $\left( 2+1\right) $
-dimensional spacetime. We define the Lagrangian density with a
Lorentz-violating interaction, where the the spacetime dimensionality is
explicitly taken into account in its definition. The work encompasses an analysis
of this model at both zero and finite-temperature, where very interesting features
are known to occur due to the spacetime dimensionality. With that in
mind we expect that the spacetime dimensionality may provide new insights
about the radiative generation of higher-derivative terms into the action, implying in a new Lorentz-violating
electrodynamics, as well the nonminimal coupling may provide interesting implications on
the thermodynamical quantities.

\end{abstract}

\thispagestyle{empty}

\newpage

\tableofcontents

\section{Introduction}

One may say that the main goal of physics is to explain phenomena in Nature, and perhaps
even to explain why physical Nature dwells in four-dimensions; however, the means that we have
come to employ in reaching this goal so far are sufficiently
intricate that it has proven useful to wander into lower-dimensional worlds, with the wishful thought
that in a simpler setting we can learn useful things about the well-recognized four-dimensional problems \cite{37}. This in fact has already happened,
initially in two-dimensions and subsequently in three-dimensions -- for instance, in condensed matter and statistical systems.

Although we live nowadays in a thrilling era of rich high-precision
experiments in particle physics, testing long-dated gauge theories, it is
matter of highly importance to improve the check upon the structural pillars
of theoretical gauge theories which may be stated as being the \textit{CPT}
theorem for Lorentz-symmetric theories. Therefore, the observation of
Lorentz or \textit{CPT} symmetry violation would be a sensitive signal for
unconventional underlying physics. This suggests that it is rather interesting to consider
theoretical mechanisms through which Lorentz or \textit{CPT} symmetry
violation might be implemented \cite{11}. In fact, there are feasible theoretical arguments that also
motivate the research on Lorentz symmetry violation, for instance one may speculate about a
deeper origin of the symmetries observed in nature, not seen as fundamental ones, but, rather, appearing
as consequences of our \textit{low-energy} world, being expected to be broken at high energies \cite{28}. Interesting studies upon
sensitive tests of Lorentz invariance in the most diverse areas of physics were subject of attention in the past decades \cite{15}.
In this spirit, numerous investigations on Lorentz-violating field theories were rekindled recently, providing a better understanding
on several aspects of this proposal \cite{19,17}.

In order to incorporate these Lorentz-violating effects in a given field theory one is led to consider
two natural ways. First: Lorentz-violating terms might
be understood in the standard model extension as vacuum expectation values of
fundamental tensor fields or, second: these Lorentz-breaking effects may be considered by
inserting new interaction terms into the Lagrangian \cite{9}, in the form of nonminimal Lorentz-violating coupling
terms. These possibilities are rather simplified
somewhat by restricting attention to operators that maintain the
conservation of energy, momentum and electric charge.
It should be emphasized that, in general, in field theory, there are two possible ways to implement the breaking of a symmetry: explicit and spontaneous. However, it has been proposed a stringent theorem in \cite{spontaneous} requiring that any breaking of Lorentz symmetry must be dynamical. On the other hand, investigations also look as cause of the breakdown of Lorentz symmetry by Nambu-Goldstone modes. Goldstone's theorem implies that the spontaneous breaking must be accompanied by massless bosons (as photons and gravitons) \cite{goldstone}.

The most natural extension to be matter of analysis would be to consider a Lorentz-violating action having the usual and simple $U\left( 1\right) $
gauge symmetry and invariance under space-time translations, so that charge,
energy and momentum are conserved (by restricting attention to the case of
constant coefficients \cite{11}). So, a natural scenario is the quantum electrodynamics (QED) with violation of Lorentz and \textit{%
CPT }symmetry; and, in fact, it have been studied intensively in the past years.
One of the most successfully studied model nowadays, in the context of Lorentz-violating theories, is the \textit{CPT}-odd
Carroll-Field-Jackiw term \cite{10}: $\epsilon _{\mu \nu \alpha \beta}\left( k_{AF}\right) ^{\mu }A^{\nu }F^{\alpha \beta }$,
several issues related to it were also topic of investigation \cite{18,5,14}.
Among the several issues analyzed lately, it is the induction of
Lorentz-violating terms by radiative corrections arising from suitable \textit{CPT}
and/or Lorentz-violating nonminimal coupling a recurrent theme in the literature; for instance, the induction of the
Chern-Simons-like action in a Lorentz-violating massless QED by a axial fermion Lorentz-violating coupling, $b_{\mu }\bar{%
\psi}\gamma ^{\mu }\gamma _{5}\psi $ \cite{5}.

Based on the previous arguments, one may start wonder the following question: what a Lorentz-violating
term, with known characteristics and properties, may induces in a field theory model if the spacetime dimensionality, where the system
is embedded, is changed, by lowering or increasing it. And, therefore, it is exactly in this thought where our inducement relies.
In this work we shall consider a particular scenario that suffices our interests, a low-dimensional field theory endowed with
Lorentz-violating coupling, to analyze mainly whether the spacetime dimensionality may account to the outcome quantities.
In fact, low-dimensional field theories were recognized, a long time ago, as
serving as laboratories where important theoretical ideas are tested in a
simple setting \cite{1,2,21,16}, specially on condensed matter systems; for instance, the quantum Hall effect \cite{27}.
Furthermore, including to that, the fact that a field theory defined in a
three-dimensional spacetime contains a highly interesting inner structure,
due to the odd spacetime dimensionality, it is rather natural to
investigate as theoretical options: how the spacetime dimensionality
accounts in the analysis upon the generation of Lorentz-violating terms, and
whether the Lorentz-violating effects changes some known theoretical
results. These two thoughts will be developed here in the context of
three-dimensional quantum electrodynamics.

First we shall discuss how the spacetime dimensionality accounts in the character of the radiative corrections, obtaining as
the outcome a perturbative generation of higher-derivative Lorentz-violating terms.\footnote{The issue of generation of
higher-derivative Chern-Simons term in three-dimensional QED is a recurrent subject in the literature \cite{30}.} Next, since it is important to analyze the
stability of these models as a function of environmental variations, we shall consider finite
temperature effects in the model by investigating whether and how the Lorentz-breaking effects change
known general properties of the form factors from the polarization tensor
\cite{21}. Actually, the issue in determining whether symmetry breaking and restoration takes place at high temperatures, where the
temperature is responsible for setting the energy scale, it was subject of analysis in diverse works \cite{5,35}.
In addition, some interesting aspects of three-dimensional Lorentz-breaking theories have been discussed recently in Ref. \cite{7}.

Within the context of the Lorentz symmetry
breaking, higher-order terms were always present in the literature: the first known
higher-derivative example is the gravitational Chern-Simons term \cite{3}; another important example is the
so-called Myers-Pospelov term \cite{12,4}. Many other analysis led to
an impressive development in the subject of generation of higher-derivative
terms \cite{13,6,24}. Originally, the presence of higher-derivatives was
known by improving the ultraviolet behavior and renormalizability of field theories, this
behavior, in particular, plays a key role in the study of the quantum theory
of gravity \cite{25}. Nevertheless, non-relativistic field theories around Lifshitz points, with anisotropic scaling between space and time, have
been of interest for a variety of problems, and have been considered as possible ultraviolet completions of
low-energy effective actions for applications to particle physics and gravity, as well as for its renormalizability,
introducing therefore unusual and interesting aspects \cite{33,35}.

In this paper, we study the aspects of the three-dimensional QED added by a nonminimal
Lorentz-violating coupling term, which takes into account the spacetime
dimensionality; it is discussed, in particular, the radiative correction at
one-loop order of the polarization tensor at both zero and finite
temperature. We start by discussing the general properties of the QED with
this nonminimal coupling term, also analyzing its relation to some other
coupling terms. We conclude the section by constructing the polarization
tensor at one-loop when considering the contributions of the nonminimal
coupling. In Sect.\ref{sec:2}, we evaluate the contribution of the
Lorentz-violating coupling, and show how it sum with the usual contribution and
to complete the discussion we write the higher-derivative contribution in
the Lagrangian level, this shows that the Maxwell-Chern-Simons electrodynamics must be modified, implying in a new Lorentz-violating
electrodynamics. In Sect.\ref{sec:3}, we present the calculation and
results of the finite temperature effects and analyze the implication of
the Lorentz-violating coupling into the known expression of the form factor
of polarization tensor. In Sect.\ref{sec:4} we summarize the results, and
present our final remarks and prospects.

\section{General discussion}

\label{sec:1}

Actually the study of higher-derivative terms in field theories for a three-dimensional spacetime is a recurrent theme in many different
context \cite{30}. Based on recent proposals regarding Lorentz-violating interaction terms in the Refs. \cite%
{13,6}, we analyze here a slightly different version of these coupling
terms. With this analysis, we expect to obtain information about the implications and effects of such known terms in a field theory,
when it is defined in a lower-dimensional spacetime; i.e., we expect to gain further insights about the role played by the spacetime dimensionality
in a Lorentz-violating field theory. Hence, we shall consider a $\left( 2+1\right) $-dimensional spacetime, and a
Lagrangian density with a Lorentz-violating interaction which takes into
account the spacetime dimensionality. Therefore, we write it as
\begin{equation}
\mathcal{L}=\mathcal{L}_{QED}+\lambda \left( K_{AF}\right) _{\mu \alpha
\beta }\bar{\psi}\gamma ^{\mu }\psi F^{\alpha \beta },  \label{eq 0.1}
\end{equation}%
based on the two-component spinor, we use a two-dimensional realization of
the Dirac algebra%
\begin{gather}
\gamma ^{0} =\sigma ^{3},\quad \gamma ^{1}=i\sigma ^{1},\quad \gamma
^{2}=i\sigma ^{2},  \label{eq 0.2} \\
\gamma ^{\mu }\gamma ^{\nu } =\eta ^{\mu \nu }-i\epsilon ^{\mu \nu \alpha
}\gamma _{\alpha },\quad \eta ^{\mu \nu }=\left( 1,-1,-1\right) ,
\end{gather}%
The $\sigma ^{a}$'s are Pauli matrices. In a three-dimensional spacetime
discrete symmetries are somewhat unusual \cite{1}. Charge conjugation,%
\begin{equation}
A^{C}_{\mu }=-A_{\mu },\quad \psi ^{C}=-i\gamma ^{1}\psi ^{\dag },
\label{eq 0.3}
\end{equation}%
leaves the equation invariant. However, under parity transformation%
\begin{align}
A_{P}^{0} &=A^{0}\left( x_{0},x^{\prime }\right) ,\quad
A_{P}^{1}=-A^{1}\left( x_{0},x^{\prime }\right) ,  \label{eq 0.4} \\
A_{P}^{2}&=A^{2}\left( x_{0},x^{\prime }\right) ,\quad \psi _{P} =-i\gamma
^{1}\psi \left( x_{0},x^{\prime }\right)  \nonumber
\end{align}%
with $x^{\prime }=\left( -x_{1},x_{2}\right)$, and time-inversion%
\begin{equation}
A_{T}^{0}=A^{0}\left( -x_{0},x\right) ,\quad A_{T}^{a}=-A^{a}\left(
-x_{0},x\right) ,  \quad \psi _{T}=-i\gamma ^{2}\psi \left( x_{0},x^{\prime }\right) ,
\label{eq 0.5}
\end{equation}%
the mass term changes its sign. But it is still \textit{CPT} invariant. The
Lorentz-violating term, though, is \textit{CPT}-odd clearly. Nevertheless, still on the Lorentz-violating term, we
may could keep the bilinear $\lambda \left( K_{F}\right) _{\mu \nu \alpha
\beta }\bar{\psi}\gamma ^{\mu }\gamma ^{\nu }\psi F^{\alpha \beta }$ as proposed in
\cite{4}, which is \textit{CPT}-even; but, due to the two-dimensional
realization, $\left[ \gamma ^{\mu },\gamma ^{\nu }\right] =-2i\epsilon ^{\mu
\nu \lambda }\gamma _{\lambda }$, we can write it equivalently as $%
-2i\lambda \left( K_{F}\right) _{\mu \nu \alpha \beta }\epsilon ^{\mu \nu
\lambda }\bar{\psi}\gamma _{\lambda }\psi F^{\alpha \beta }$, and finally
relate the coefficients as: $\left( K_{AF}\right) _{\lambda \alpha \beta
}=-2i\lambda \left( K_{F}\right) _{\quad \alpha \beta }^{\mu \nu }\epsilon
_{\mu \nu \lambda }$.\footnote{%
This can be seen as related also with the four-dimensional term in \cite{4},
with the nonminimal coupling given by: $-g\epsilon_{ \mu \nu \lambda \rho
}b^{\rho }\bar{\psi}\gamma ^{\mu }\psi F^{\nu \lambda }$, where the $b^{\rho
}$ is a constant vector implementing the Lorentz symmetry breaking. This
interaction is known to generate radiatively the Myers-Pospelov term and a
higher-derivative Carroll-Field-Jackiw term.} This shows, therefore, that both
choices to the nonminimal coupling are related. Thus, we will
investigate here the first case as proposed above in the Eq.\eqref{eq 0.1}.

Furthermore, the tensor $\left( K_{AF}\right) _{\mu \alpha \beta }$ in \eqref{eq 0.1} is
antisymmetric in the last two indices: $\left( K_{AF}\right) _{\mu \alpha
\beta }=\left( K_{AF}\right) _{\mu \left[ \alpha \beta \right] }$. Note
that, bearing that in mind, we may parameterize the tensor $\left( K_{AF}\right) _{\mu \alpha \beta }$
by the vector defined as: $\kappa _{\alpha }=\left( K_{AF}\right) ^{\mu}_{~~ \alpha \mu }$, which fulfills the following linear
combination
\begin{equation}
\left( K_{AF}\right) _{\mu \alpha \beta }=\frac{1}{2}\left( \eta _{\mu \beta
}\kappa _{\alpha }-\eta _{\mu \alpha }\kappa _{\beta }\right) .
\label{eq 0.6}
\end{equation}%
With this particular choice one finds%
\begin{equation}
\mathcal{L}=\bar{\psi}\left( i\gamma .\partial -m\right) \psi -\frac{1}{4}%
F_{\mu \nu }F^{\mu \nu }-\frac{1}{2\xi }\left( \partial _{\mu }A^{\mu
}\right) ^{2}+\mathcal{L}_{int},  \label{eq 0.8}
\end{equation}%
with the interaction term given by%
\begin{equation}
\mathcal{L}_{int}=-gA_{\mu }\bar{\psi}\gamma ^{\mu }\psi +\lambda \kappa
_{\alpha }\bar{\psi}\gamma _{\mu }\psi F^{\alpha \mu }.  \label{eq 0.9}
\end{equation}%
One should notice that the minimal coupling $g$ and the nonminimal one,
proportional to $\lambda $, have different length dimensions \cite{31}. At last we can
derive the Feynman rules for the perturbative treatment,
\begin{equation}
\Lambda ^{\mu }\left( p,q;r\right) =\left[ ig\gamma ^{\mu }-\lambda \kappa
_{\sigma }\gamma _{\alpha }\left( r^{\alpha }\eta ^{\mu \sigma }-r^{\sigma
}\eta ^{\mu \alpha }\right) \right] \delta \left( p+q+r\right) .
\label{eq 0.11}
\end{equation}%
Through that we can write the polarization tensor at one-loop order as%
\begin{equation}
\Pi ^{\mu \nu }\left( k\right) =i\int \frac{d^{\omega }p}{\left( 2\pi
\right) ^{\omega }}\text{Tr}\bigg[ \Lambda ^{\mu }\left( p,-p-k;k\right)
S\left( p+k\right)  \Lambda ^{\nu }\left( p+k,-p;-k\right) S\left( p\right) \bigg] ,
\label{eq 0.12}
\end{equation}%
Such expression takes into account all contributions for the radiative
corrections from the Lorentz-violating term, including the
usual $QED_{3}$ as well. As it will be shown next, the way that the relevant
contributions sum it will be rather interesting and surprisingly particular.
Now we will investigate Eq.\eqref{eq 0.12} at $T=0$ and at $T\neq 0$ to see
which implications may be generated at one-loop approximation.


\section{One-loop correction at $T=0$}

\label{sec:2}

Nevertheless, we start by evaluating Eq.\eqref{eq 0.12} at $T=0$ through dimensional regularization. Therefore,%
\begin{align}
\Pi ^{\mu \nu }\left( k\right) =&-i\mu ^{3-\omega }\int \frac{d^{\omega }p}{%
\left( 2\pi \right) ^{\omega }} \text{Tr}\bigg[\left( ig\gamma ^{\mu }-\lambda \gamma ^{\alpha
}\left( \kappa ^{\mu }k_{\alpha }-\eta _{\alpha }^{\mu }\kappa _{\beta
}k^{\beta }\right) \right) \frac{\left( -i\right) }{\gamma .\left(
p+k\right) -m} \notag \\
&\times \left( ig\gamma ^{\nu }-\lambda \gamma ^{\sigma }\left( \kappa
^{\nu }k_{\sigma }-\eta _{\sigma }^{\nu }\kappa _{\rho }k^{\rho }\right)
\right) \frac{\left( -i\right) }{\gamma .p-m}\bigg].
\end{align}%
Actually, due to the linearity of $\Lambda $ regarding its dependence to the
$\gamma $'s, we can evaluate the trace operation in general terms as%
\begin{equation}
N^{\mu \nu }\left( p,k\right) =\text{Tr}\left[ \gamma ^{\mu }\left( \gamma
.\left( p+k\right) +m\right) \gamma ^{\nu }\left( \gamma .p+m\right) \right],
\end{equation}%
it suffices the following three-dimensional result (for two-components
realization)%
\begin{equation}
\gamma ^{\mu }\gamma ^{\nu }=\eta ^{\mu \nu }-i\epsilon ^{\mu \nu \alpha
}\gamma _{\alpha },
\end{equation}%
resulting, thus, into the expression\footnote{%
As we do not have the presence of dimension-dependent $\gamma _{5}$ matrix here
there are not ambiguities in this calculation at dimensional regularization.}%
\begin{equation}
N^{\mu \nu }\left( p,k\right) =2\bigg\{ 2p^{\mu }p^{\nu }+k^{\mu }p^{\nu
}+k^{\nu }p^{\mu }+im\epsilon ^{\mu \nu \delta }k_{\delta } +\left( m^{2}-p.\left( p+k\right) \right) \eta ^{\mu \nu }\bigg\} .
\label{eq 1.3}
\end{equation}%
First we calculate%
\begin{equation}
\Pi _{\left( a\right) }^{\mu \nu }\left( k\right) =-ig^{2}\mu ^{3-\omega
}\int \frac{d^{\omega }p}{\left( 2\pi \right) ^{\omega }}\frac{N^{\mu \nu
}\left( p,k\right) }{\left( \left( p+k\right) ^{2}-m^{2}\right) \left(
p^{2}-m^{2}\right) },
\end{equation}%
now, following standard rules of dimensional regularization, it is not
complicated to one finds%
\begin{equation}
\Pi _{\left( a\right) }^{\mu \nu }\left( k\right) =\frac{g^{2}}{2\pi }%
\left( k^{2}\eta ^{\mu \nu }-k^{\mu }k^{\nu }\right) f_{2}\left(
k,m\right)  +\frac{g^{2}}{4\pi }im\epsilon ^{\mu \nu \delta }k_{\delta }f_{1}\left(
k,m\right) ,  \label{eq 1.4}
\end{equation}%
where we have defined the functions
\begin{align}
f_{1}\left( k,m\right) &=\int_{0}^{1}dz\frac{1}{\left(
m^{2}-z\left( 1-z\right) k^{2}\right) ^{\frac{1}{2}}}, \\
f_{2}\left( k,m\right)&=\int_{0}^{1}dz\frac{z\left( 1-z\right) }{%
\left( m^{2}-z\left( 1-z\right) k^{2}\right) ^{\frac{1}{2}}}.  \label{eq 1.5}
\end{align}
Moreover, after some algebraic manipulation, we can calculate the next
contribution%
\begin{equation}
\Pi _{\left( b\right) }^{\mu \nu }\left( k\right) +\Pi _{\left( c\right)
}^{\mu \nu }\left( k\right) =g\lambda \mu ^{3-\omega } \int \frac{d^{\omega
}p}{\left( 2\pi \right) ^{\omega }} \frac{\left( \kappa ^{\nu }k_{\sigma }N^{\mu \sigma
}+\kappa^{\mu}k_{\alpha}N^{\alpha\nu }-2\kappa _{\rho }k^{\rho }N^{\mu \nu
}\right) }{\left( \left( p+k\right) ^{2}-m^{2}\right) \left(
p^{2}-m^{2}\right) } .
\end{equation}%
Again, through well-known methods, one gets%
\begin{align}
\Pi _{\left( b\right) }^{\mu \nu }\left( k\right) +\Pi _{\left( c\right)
}^{\mu \nu }\left( k\right) =&\frac{g\lambda }{2\pi }m\epsilon ^{\mu \nu
\delta }k_{\delta }\left( \kappa _{\rho }k^{\rho }\right) f_{1}\left(
k,m\right) \notag \\
&+ i\frac{g\lambda }{\pi }\left( k^{\mu }k^{\nu }-k^{2}\eta ^{\mu \nu
}\right) \left( \kappa _{\rho }k^{\rho }\right) f_{2}\left(
k,m\right).  \label{eq 1.6}
\end{align}
Actually, it is related to the first contribution as%
\begin{equation}
\Pi _{\left( b\right) }^{\mu \nu }\left( k\right) +\Pi _{\left( c\right)
}^{\mu \nu }\left( k\right) =-i\frac{2\lambda }{g}\left( \kappa _{\rho
}k^{\rho }\right) \Pi _{\left( a\right) }^{\mu \nu }\left( k\right) .
\end{equation}%
At last, we compute the fourth contribution,%
\begin{align}
\Pi _{\left( d\right) }^{\mu \nu }\left( k\right) =&i\lambda ^{2}\mu
^{3-\omega }\left( \kappa ^{\mu }k_{\alpha }-\eta _{\alpha }^{\mu }\kappa
_{\beta }k^{\beta }\right) \left( \kappa ^{\nu }k_{\sigma }-\eta _{\sigma
}^{\nu }\kappa _{\rho }k^{\rho }\right)  \nonumber \\
& \times \int \frac{d^{\omega }p}{\left( 2\pi \right) ^{\omega }}\frac{%
N^{\alpha \sigma }\left( p,k\right) }{\left( \left( p+k\right)
^{2}-m^{2}\right) \left( p^{2}-m^{2}\right) },
\end{align}%
which results simply into%
\begin{align}
\Pi _{\left( d\right) }^{\mu \nu }\left( k\right)=\frac{\lambda ^{2}}{%
2\pi }\left(\kappa _{\beta }k^{\beta }\right)^{2}\left( k^{\mu }k^{\nu
}-k^{2}\eta ^{\mu \nu }\right) f_{2}\left( k,m\right)  -\frac{\lambda ^{2}}{4\pi }\left(\kappa _{\beta }k^{\beta
}\right)^{2}\epsilon ^{\mu \nu \delta }k_{\delta }imf_{1}\left(
k,m\right) .  \label{eq 1.7}
\end{align}%
We also can write it as%
\begin{equation}
\Pi _{\left( d\right) }^{\mu \nu }\left( k\right) =-\frac{\lambda ^{2}}{g^{2}%
}\left(\kappa _{\beta }k^{\beta }\right)^{2}\Pi _{\left( a\right) }^{\mu \nu
}\left( k\right) .
\end{equation}%
Therefore, by summing the four contributions one determines%
\begin{align}
\Pi ^{\mu \nu }\left( k\right) =&\left[1-i\frac{2\lambda }{g}\left( \kappa
_{\rho }k^{\rho }\right) -\frac{\lambda ^{2}}{g^{2}}\left(\kappa _{\sigma
}k^{\sigma }\right)^{2}\right]\Pi _{\left( a\right) }^{\mu \nu }\left(
k\right) ,  \nonumber \\
=&\left( 1-i\frac{\lambda }{g}\left( \kappa _{\rho }k^{\rho }\right)
\right) ^{2}\Pi _{\left( a\right) }^{\mu \nu }\left( k\right) ,
\label{eq 1.8}
\end{align}%
where $\Pi _{\left( a\right) }^{\mu \nu }\left( k\right) $ is given by %
\eqref{eq 1.4}. Besides that, it follows from \eqref{eq 1.8} the verification
that%
\begin{equation}
k_{\mu }\Pi ^{\mu \nu }\left( k\right) =0.
\end{equation}%
Assuring, therefore, that at 1-loop level, the modified Lagrangian
\eqref{eq 0.8} is free of gauge anomalies. The expressions $f_{1}\left(
k,m\right) $ and $f_{2}\left( k,m\right) $ are explicitly
evaluated for $0<k^{2}<4m^{2}$%
\begin{align}
f_{1}\left( k,m\right) =&\frac{2}{\sqrt{k^{2}}}\coth ^{-1}\left(
\frac{2m}{\sqrt{k^{2}}}\right) , \\
f_{2}\left( k,m\right) =&-\frac{\left\vert m\right\vert }{2k^{2}}+%
\frac{\left( 1+\frac{4m^{2}}{k^{2}}\right) }{4\sqrt{k^{2}}}\coth ^{-1}\left(
\frac{2m}{\sqrt{k^{2}}}\right) ,  \label{eq 1.9}
\end{align}%
while, by analytic continuation, one finds the expressions for $k^{2}>4m^{2}$%
. Moreover, a limit of interest is for $\frac{k^{2}}{m^{2}}\rightarrow 0$, where
the low-energy effects $\left( \text{in which }k^{2}\ll m^{2}\right) $ take
place,%
\begin{equation}
f_{1}\left( k,m\right) =\frac{1}{\left\vert m\right\vert },\quad
f_{2}\left( k,m\right) =\frac{1}{6\left\vert m\right\vert }.
\label{eq 1.10}
\end{equation}%
One finds, thus, that the mixed contribution \eqref{eq 1.6} yields the
Lagrangian,%
\begin{equation}
\mathcal{L}_{\left( b,c\right) }=\frac{g\lambda }{12\pi \left\vert
m\right\vert }\kappa _{\rho }F_{\mu \nu }\partial ^{\rho }F^{\mu \nu }+\frac{%
g\lambda }{4\pi }\frac{m}{\left\vert m\right\vert }\kappa ^{\rho }\epsilon
^{\mu \nu \delta }A_{\mu }\partial _{\rho }F_{\nu \delta },  \label{eq 1.12}
\end{equation}%
whereas, for the full Lorentz-violating contribution \eqref{eq 1.7} it
follows%
\begin{equation}
\mathcal{L}_{\left( d\right) }=\frac{\lambda ^{2}}{24\pi \left\vert
m\right\vert }F_{\mu \nu }\left(\kappa.\partial \right)^{2} F^{\mu \nu }+%
\frac{\lambda ^{2}}{4\pi }\frac{m}{\left\vert m\right\vert }A_{\mu }\epsilon
^{\mu \nu \delta }\left(\kappa.\partial \right)^{2}F_{\nu \delta },
\label{eq 1.13}
\end{equation}%
in coordinate space. Although all contributions can also be rewritten as%
\begin{equation}
\mathcal{L}_{t }=\frac{g^{2}}{24\pi \left\vert
m\right\vert }F_{\mu \nu }\left( 1+\frac{\lambda }{g}\left(
\kappa.\partial\right) \right) ^{2}F^{\mu \nu } +\frac{g^{2}}{4\pi }\frac{m}{\left\vert m\right\vert }A_{\mu }\left( 1+%
\frac{\lambda }{g}\left( \kappa .\partial \right) \right) ^{2}\epsilon ^{\mu
\nu \delta }F_{\nu \delta }. \label{eq 1.13c}
\end{equation}
As present here in the Eq.\eqref{eq 1.13c}, higher-derivative terms in three-dimensional field theories
is a theme analyzed recurrently in the literature \cite{30}. \footnote{Moreover, a related situation also appears
for the case of spontaneous parity breaking \cite{1}.}
Furthermore, our analysis has shown that the radiative
correction to the one-loop vacuum polarization had lead to a higher-order dimension-three CPT-odd term \eqref{eq 1.13c},
by the addition of a nonminimal coupling. This shows that the
Maxwell-Chern-Simons electrodynamics must be modified by the inclusion of such term in its structure, implying in a new Lorentz-violating
electrodynamics. Looking
up to the second premise of the present work, we shall discuss whether the
Lorentz-violating effects changes some known theoretical results. To
accomplish this analysis we will discuss how the Lorentz symmetry
breaking may accounts to some known results of thermal effects involving the
form factors of the polarization tensor.


\section{One-loop correction at $T\neq 0$}

\label{sec:3}

As discussed earlier it has become an increasingly important subject of
analysis nowadays field theory models endowed with Lorentz-breaking terms,
and it is equally important to address the question of stability of these
models as a function of environmental variations. Adding to that the fact
that the three-dimensional QED at finite temperature is known to possesses several interesting
properties \cite{21} -- observables are finite and the theory is well
defined,-- and also the interesting result of the previous section, it is
expected to obtain here further insights about whether Lorentz-breaking effects
may contribute to the well-known properties of $QED_{3}$. \footnote{Actually, studies on Lorentz-violating effects at finite temperature, but in a four-dimensional field theory, was presented in \cite{finitemp}.} It
is worth of stressing, however, that when we evaluate the electric
and magnetic screening masses, we will present the resulting expression in
the massless limit in order to have a much simpler but still elucidating
content.

On its most general covariant form the polarization tensor involves a
dependence on $\eta ^{\mu \nu }$, external momentum $k^{\mu }$ and at the
thermal bath velocity $u^{\mu }$ \cite{22}. Moreover, due to the
Ward-Takahashi identity we know that the self-energy tensor must be
transverse to external momentum,%
\begin{equation}
k_{\mu }\Pi ^{\mu \nu }\left( k\right) =0.  \label{eq 2.1}
\end{equation}%
This identity works as a guideline to us to construct, the most general
covariant structure for $\Pi ^{\mu \nu }$ subject to a thermal bath, it
would be%
\begin{equation}
\Pi ^{\mu \nu }\left( k\right) =P^{\mu \nu }\Pi _{S}\left( k^{2}\right) +%
\frac{u^{\mu }u^{\nu }}{u^{2}}\Pi _{u}\left( k^{2}\right) +i\epsilon ^{\mu
\nu \alpha }k_{\alpha }\Pi _{A}\left( k^{2}\right) ,  \label{eq 2.2}
\end{equation}%
where,%
\begin{equation}
P_{\mu \nu }=\eta _{\mu \nu }-\frac{k_{\mu }k_{\nu }}{k^{2}}-\frac{u_{\mu
}u_{\nu }}{u^{2}}.
\end{equation}%
We shall evaluate the self-energy at the static limit (analytic
expressions), $k_{0}\rightarrow 0$, and it becomes interesting to choose the
thermal bath as at rest, i.e., $u^{\mu }=n^{\mu }$; where $n^{\mu }=\left( 1,%
\overrightarrow{0}\right) $. All of that allows one to obtain
the form factors,%
\begin{align}
\Pi _{u}\left( k^{2}\right) =&n_{\mu }n_{\nu }\Pi ^{\mu \nu },
\label{eq 2.3a} \\
\Pi _{A}\left( k^{2}\right) =&\frac{1}{2ik^{2}}\epsilon _{\mu \nu \sigma
}k^{\sigma }\Pi ^{\mu \nu },  \label{eq 2.3b} \\
\Pi _{S}\left( k^{2}\right) =&\eta _{\mu \nu }\Pi ^{\mu \nu }-\Pi
_{u}\left( k^{2}\right) .  \label{eq 2.3c}
\end{align}%
We shall evaluate Eq.\eqref{eq 0.12} at $T\neq 0$ through
imaginary-time formalism \cite{22}. First, remembering that in the
imaginary-time approach the fermionic momentum is replaced by $%
p_{0}\rightarrow i\omega _{l}=\frac{\pi i}{\beta }\left( 2l+1\right) $,
while the external momentum is modified as $k_{\mu }=\left( k_{0},%
\overrightarrow{k}\right) \rightarrow \left( i\omega _{n},\overrightarrow{k}%
\right) $, where $\omega _{n}=\frac{2\pi n}{\beta }$ is the bosonic
Matsubara frequency.

The momentum sum will be evaluated at the static limit, where the self-energy
possesses an analytic structure; while the momentum integral is
evaluated at the hard thermal loop, where all the external momenta are
neglected with the internal one \cite{22}. Therefore, these information lead to: $
\Pi ^{\mu \nu }\left( k\right) =\Pi ^{\mu \nu }\left( k_{0},\overrightarrow{k%
}\right) \rightarrow \Pi ^{\mu \nu }\left( 0,\overrightarrow{k}\right)$,
also the numerator $N^{\mu \nu }\left( p\right) $, Eq.\eqref{eq 1.3},
containing the $\gamma $-matrices, reduces to%
\begin{equation}
N^{\mu \nu }\left( p\right) =2\left\{ 2p^{\mu }p^{\nu }+im\epsilon ^{\mu \nu
\delta }k_{\delta }+\left( m^{2}-p^{2}\right) \eta ^{\mu \nu }\right\} .
\end{equation}%
With the above results we are able to evaluate the sum over the
fermionic frequencies remaining, therefore, only the momentum integral.
Thus, we can write the four contribution as%
\begin{align}
\Pi _{\left( a\right) }^{\mu \nu }\left( 0,\overrightarrow{k}\rightarrow
0\right)  =&-g^{2}\tilde{\Pi}^{\mu \nu }, \\
\left(\Pi _{\left( b\right) }^{\mu \nu }+ \Pi _{\left( c\right) }^{\mu \nu }\right)\left( 0,\overrightarrow{k}\rightarrow
0\right)  =&g\lambda \left( \kappa ^{\mu }k_{\alpha }\tilde{\Pi}^{\alpha \nu }
+\kappa ^{\nu }k_{\sigma }\tilde{\Pi}^{\mu \sigma }-2\left(\kappa .k \right)\tilde{\Pi}^{\mu \nu }\right) , \\
\Pi _{\left( d\right) }^{\mu \nu }\left( 0,\overrightarrow{k}\rightarrow
0\right)  =&-\lambda ^{2}\left( \kappa ^{\mu }k_{\alpha }-\eta _{\alpha
}^{\mu }\kappa _{\beta }k^{\beta }\right) \left( \kappa ^{\nu }k_{\sigma
}-\eta _{\sigma }^{\nu }\kappa _{\rho }k^{\rho }\right) \tilde{\Pi}^{\alpha
\sigma }.
\end{align}%
where we have defined the quantity%
\begin{align}
\tilde{\Pi}^{\mu \nu }\left( 0,\overrightarrow{k}\rightarrow 0\right)=&2\int
\frac{d^{3}pp^{\mu }p^{\nu }}{\left( 2\pi \right) ^{2}}I^{\left( 2\right)
}\delta \left( p_{0}-\omega _{p}\right)  +\left( \eta ^{\mu \nu }-2n^{\mu
}n^{\nu }\right) \int \frac{d^{2}p}{\left( 2\pi \right) ^{2}}I^{\left(
1\right) } \notag \\
& +im\epsilon ^{\mu \nu \delta }k_{\delta }\int \frac{d^{2}p}{\left(
2\pi \right) ^{2}}I^{\left( 2\right) },
\end{align}
and introduced these as well%
\begin{align}
I^{\left( 1\right) }\left( 0,\overrightarrow{k}\rightarrow 0\right) =&-\frac{1%
}{\omega _{p}}n_{F}\left( \omega _{p}\right) ,  \\
I^{\left( 2\right) }\left( 0,\overrightarrow{k}\rightarrow 0\right) =&\frac{1}{2\omega
_{p}^{2}}n_{F}^{\prime }\left( \omega _{p}\right)-\frac{1%
}{2\omega _{p}{}^{3}}n_{F}\left( \omega _{p}\right) .  \label{eq 2.12}
\end{align}%
where $n_{F}\left( \theta \right) =\frac{1}{1+e^{\beta \theta }}$ is the fermionic distribution.
Thus, we start by computing $\Pi _{u}\left( k^{2}\right) $, from Eq.%
\eqref{eq 2.3a} one finds%
\begin{align}
\Pi _{u}\left( k^{2}\right) =&-\left( g^{2}+2g\lambda \kappa _{\rho
}k^{\rho }+\lambda ^{2}\left( \kappa _{\rho }k^{\rho }\right) ^{2}\right)
\tilde{\Pi}^{00} -\lambda ^{2}\kappa ^{0}\kappa ^{0}k_{\alpha }k_{\sigma }\tilde{\Pi}%
^{\alpha \sigma } \nonumber \\
&+\lambda \left( g+\lambda \left( \kappa _{\rho }k^{\rho }\right) \right)
\kappa ^{0}k_{\sigma }\left( \tilde{\Pi}^{\sigma 0}+\tilde{\Pi}^{0\sigma
}\right) .  \label{eq 2.16}
\end{align}%
We will keep considering the general case $k=\left( 0,\overrightarrow{k}%
\right) $. Let us evaluate the above terms directly. It is known: $%
d^{2}p=2pdpd\psi $, where we choose $\psi $ as the angle between $p$ and $k$%
. Therefore, one obtains%
\begin{equation}
\tilde{\Pi}^{00}\left( 0,\overrightarrow{k}^{2}=m_{mag}^{2}%
\right) =-\frac{1}{%
2\pi }\frac{1}{\beta }\ln \left[ 1+e^{m\beta }\right] +\frac{m}{2\pi }\frac{1}{1+e^{-m\beta }}\equiv -\tilde{m}_{el}^{2},
\label{eq 2.17}
\end{equation}%
a relation that defines the usual electric screening mass. In the massless limit the
electric mass goes as: $ \tilde{m}_{el}^{2}=\frac{1}{2\pi }\frac{1}{\beta }\ln 2$. Discarding the independent temperature terms, one determines the complete
longitudinal contribution%
\begin{align}
\Pi _{u}\left( 0,\overrightarrow{k}\rightarrow 0\right)=&\frac{1}{2\pi }%
\left( g^{2}+2g\lambda \left( \kappa .k \right) +\lambda
^{2}\left( \kappa .k \right) ^{2}\right) \left[ \frac{1}{\beta
}\ln \left[ 1+e^{m\beta }\right] -\frac{m}{1+e^{-m\beta }}\right] \notag \\
&+\frac{\lambda ^{2}}{4\pi }\left(\kappa .n\right)^{2}k^{2}\bigg[ \frac{4}{\beta }\ln \left[ 1+e^{m\beta }\right] -\frac{m}{%
1+e^{-m\beta }}+m\tanh \left( \frac{m\beta }{2}\right) \bigg] . \label{eq 2.53}
\end{align}
In the massless limit the complete screening electric mass reads as%
\begin{equation}
m_{el}^{2}=\frac{1}{2\pi \beta }\frac{\left( g+\lambda \left( \kappa
.k\right) \right) ^{2}}{\left( 1-\frac{\lambda ^{2}}{\pi \beta }\left(
\kappa .n\right) ^{2}\ln 2\right) }\ln 2.
\end{equation}%
Of course this screening mass regulates infrared divergences for the
longitudinal portion of the photon propagator. However, it still necessary to check
whether this is true for the transverse $\left( \Pi _{S}\right) $ and
Chern-Simons $\left( \Pi _{A}\right) $ portions.

Next, we compute the Chern-Simons form factor through Eq.\eqref{eq 2.3b}.
After some integral evaluation and algebraic manipulation we have%
\begin{equation}
\Pi _{A}\left( 0,\overrightarrow{k}\rightarrow 0\right) =-\frac{1}{4\pi }%
\left( g+\lambda \left( \kappa .k\right) \right) ^{2}\tanh \left( \frac{%
m\beta }{2}\right) .  \label{eq 2.23}
\end{equation}%
One may easily check that the zero temperature limit correctly recovers Eq.%
\eqref{eq 1.4}; but, on the other hand, as the fermion mass goes to zero, Eq.%
\eqref{eq 1.4} provides a finite result whereas the finite temperature
expression \eqref{eq 2.23} approaches zero if $T$ is kept finite (the
correct result is obtained if one takes the limit before performing the
integral).

Finally, we compute the last form factor: $\Pi _{S}\left(
k^{2}\right) $ through Eq.\eqref{eq 2.3c}. As we already have evaluated $\Pi
_{u}$ in Eq.\eqref{eq 2.53}, it only remains to compute the quantity $\eta
_{\mu \nu }\Pi ^{\mu \nu }$. Thus, evaluating the momentum integral, one gets
\begin{align}
\eta _{\mu \nu }\Pi ^{\mu \nu }\left( 0,\overrightarrow{k}\rightarrow
0\right)  =&-\frac{m}{2\pi }\left( \left( g+\lambda \left( \kappa .k\right)
\right) ^{2}-\frac{1}{2}\lambda ^{2}\left( \kappa .\kappa \right)
k^{2}\right) \tanh \left( \frac{m\beta }{2}\right)   \nonumber \\
&+\frac{1}{2\pi \beta }\left( g^{2}+2\lambda ^{2}\left( \kappa .\kappa
\right) k^{2}-\lambda ^{2}\left( \kappa .k\right) ^{2}\right) \ln \left[
1+e^{m\beta }\right] \notag \\
& -\frac{\lambda ^{2}}{4\pi }\left( \kappa .\kappa \right) k^{2}\frac{m}{%
1+e^{-m\beta }}.  \label{eq 2.28}
\end{align}%
Therefore, at the massless limit, one finds for the magnetic screening mass
\begin{align}%
m_{mag}^{2}\equiv &-\Pi _{S}\left( 0,\overrightarrow{k}^{2}=m_{mag}^{2}%
\right)\notag \\
& =\frac{1}{\pi \beta }\bigg(g\lambda \left( \kappa .k\right) +\lambda
^{2}\left( \kappa .k\right) ^{2}+m_{mag}^{2}\lambda ^{2}\left( \kappa
.n\right) ^{2}-m_{mag}^{2}\lambda ^{2}\left( \kappa .\kappa \right) \bigg)%
\ln 2,
\end{align}%
or
\begin{equation}
m_{mag}^{2}=\frac{1}{\pi \beta }\frac{\left( g\lambda \left( \kappa
.k\right) +\lambda ^{2}\left( \kappa .k\right) ^{2}\right) }{\left( 1-\frac{%
\lambda ^{2}}{\pi \beta }\left( \kappa .n\right) ^{2}\ln 2+\frac{\lambda ^{2}%
}{\pi \beta }\left( \kappa .\kappa \right) \ln 2\right) }\ln 2. \label{eq 2.28a}
\end{equation}%
It is easy to see that the usual QED contribution of magnetic screening
mass is null, and it remains true to all orders \cite%
{21}; nevertheless, there are new contributions arising from the
Lorentz-violating interaction, generating therefore a nontrivial magnetic
mass. A similar thermal effect is also present in the noncommutative QED
\cite{23}. In fact, it should be emphasized that the previous result has
an important implication in the fermionic self-energy. It is well known that a vanishing magnetic mass in QED
implies that the fermionic self-energy is infrared divergent \cite{21} (actually, one may relate this divergence
to the presence of $1/q^2$ terms in the complete expression for the photon propagator). But,
as we have showed right above, in the Eq.\eqref{eq 2.28a}, we have a nonvanishing
magnetic that changes the propagator pole as: $(q^2-m_{mag}^{2})^{-1}$, thus the logarithmic infrared divergence
is absent from the fermion propagator in our present case.


\section{Concluding Remarks}

\label{sec:4}

In this work we developed a study in a modified quantum
electrodynamics endowed with a Lorentz-violating nonminimal coupling, embedded in a three-dimensional spacetime. The main aim of the paper consisted
in analyze whether the spacetime dimensionality accounts to the Lorentz-violating effects in the outcome quantities. The analyzed
nonminimal coupling takes into account the spacetime dimensionality;
moreover, due to our choice of two-component realization for the spinor
field, it was shown how this coupling is related to other couplings known
in the literature. The present study was mainly motivated by the numerous
theoretical ideas that were tested in the simple setting of
low-dimensional field theories, in particular in condensate matter systems, e.g., the quantum Hall effect. And, in our point of view, the behavior of
Lorentz-violating couplings certainly is one of the ideas that deserves a detailed
study in this scenario \cite{7}. Hence, it is worth of investigation as
theoretical options: how the spacetime dimensionality accounts in the
analysis upon the generation of Lorentz-violating terms, and whether the
Lorentz-violating effects change some known theoretical results. These two
thoughts guided us in the development of the present analysis. In order to
accomplish that, we first discussed how the spacetime dimensionality
accounts in the perturbative generation of higher-derivative
Lorentz-violating terms by evaluating explicitly the nonminimal coupling
contribution to the polarization tensor and how it sums to the usual one.
Moreover, such contribution was written in the context of Lagrangian
density to elucidate its content. This showed that the
Maxwell-Chern-Simons electrodynamics must be modified by the inclusion of such term in its structure, implying in a new Lorentz-violating
electrodynamics. Next, we considered the role played by thermal effects in the model, investigating how the Lorentz-breaking effects
change the known general properties of the form factors from the
polarization tensor. The most interesting result on the thermal analysis was the generation of a
nontrivial magnetic screening mass arising from the Lorentz-violating
coupling; moreover, this nonvanishing screening magnetic mass
also results in a finite infrared expression for the fermionic self-energy \cite{21}. A similar thermal effect is also found in the noncommutative QED \cite{23}.

\section*{Acknowledgment}

The author would like to thanks the anonymous referee for his/her
comments and suggestions to improve this paper. RB thanks FAPESP for full support and Prof. B.M. Pimentel for his fruitful criticism and comments.

\bibliographystyle{elsarticle-num}

\end{document}